\documentclass[aps,pra,reprint,superscriptaddress,amsmath,amssymb]{revtex4-1}
\pdfoutput=1
\usepackage{graphicx}
\usepackage{bbm} 

\usepackage{hyperref}
\hypersetup{pdftitle={Tunneling in the self-trapped regime of a two-well BEC},
pdfauthor={Tadeusz Pudlik, Holger Hennig, Dirk Witthaut, David K. Campbell}}

\usepackage[usenames,dvipsnames]{xcolor}

\newcommand{\indicator}[1]{\mathbbm{1}\left( #1 \right)}
\newcommand{\arccosh}{\mathrm{arccosh\,}}
\newcommand{\E}{\mathrm{E}}
\newcommand{\K}{\mathrm{K}}

\newcommand{\ket}[1]{\left| #1 \right>}
\newcommand{\melement}[3]{\left< #1 \right| #2 \left| #3 \right>}
\newcommand{\unit}[1]{\,\mathrm{#1}}
\newcommand{\abs}[1]{\left|#1\right|}

\newcommand{\refeq}[1]{(\ref{#1})}

\begin{document}
\title{Tunneling in the self-trapped regime of a two-well BEC}

\author{Tadeusz Pudlik}
\affiliation{Department of Physics, Boston University, Boston, MA 02215, USA}
\author{Holger Hennig}
\affiliation{Department of Physics, Harvard University, Cambridge, MA 02138, USA}
\author{Dirk Witthaut}
\affiliation{Network Dynamics, Max Planck Institute for Dynamics and Self-Organization (MPIDS), 37077 G\"ottingen, Germany}
\affiliation{Forschungszentrum J\"ulich, Institute of Energy and Climate Research -- Systems Analysis and Technology Evaluation (IEK-STE), 52425 J\"ulich, Germany}
\affiliation{Institute for Theoretical Physics, University of Cologne, 50937 K\"oln, Germany}
\author{David K.~Campbell}
\email[To whom correspondence should be addressed, at ]{dkcampbe@bu.edu}
\affiliation{Department of Physics, Boston University, Boston, MA 02215, USA}

\date{\today}

\begin{abstract}
Starting from a mean-field model of the Bose-Einstein condensate dimer, we reintroduce classically forbidden tunneling through a Bohr-Sommerfeld quantization approach.  We find closed-form approximations to the tunneling frequency more accurate than those previously obtained using different techniques.  We discuss the central role that tunneling in the self-trapped regime plays in a quantitatively accurate model of a dissipative dimer leaking atoms to the environment.  Finally, we describe the prospects of experimental observation of tunneling in the self-trapped regime, both with and without dissipation.
\end{abstract}
\maketitle

A Bose-Einstein condensate of atoms in two modes (BEC dimer) is a simple interacting quantum system that has recently become accessible to increasingly precise experiments~\cite{Gati2007}.  It has been used to demonstrate matter-wave interferometry~\cite{Schumm2005}, number squeezing~\cite{Jo2007, Esteve2008, Gross2010} and measurements transcending the standard quantum limit~\cite{Cadoret2009,Lucke2011}, and its prospective applications include gravity detectors~\cite{Hall2007}, noise thermometers~\cite{Gati2006} and tests of the EPR paradox~\cite{He2011}.  Especially exciting is the opportunity to study the gradual emergence of classical mechanics as the number of particles in the system is increased~\cite{Zibold2010}.

The simplest theoretical approach to the BEC dimer is the two mode mean-field model~\cite{Smerzi1997,Raghavan1999}.  In this model, a phenomenon known as self-trapping takes place: a coherent state prepared in certain regions of phase space remains in the neighborhood of the nearest stable fixed point (self-trapping point) forever.  Self-trapping has been experimentally observed for relatively short times~\cite{Albiez2005,Zibold2010}.  In the quantum treatment of the two-mode model, tunneling between the two self-trapping fixed points eventually occurs.  This process of ``quantum sloshing'' generates macroscopic entanglement between the two wells of the dimer~\cite{Carr2010}.  The time scale on which tunneling takes place can be found numerically by directly integrating the Schr\"odinger equation, but this offers little insight into the process.  An analytical estimate of the tunneling frequency has been obtained using quantum perturbation theory (see~\cite{Salgueiro2007} and references therein), but the expansion employed is only valid in a parameter range where the tunneling frequency is exponentially small.

In this paper, we use the semiclassical quantization approach to the Bose-Hubbard dimer pioneered in~\cite{Graefe2007} to obtain highly accurate analytical approximations to the tunneling frequency of the two lowest-energy states.  Unlike quantum perturbation theory, the semiclassical techniques remain applicable as long as approximately self-localized quantum states exist.  This showcases the power of the semiclassical approach to many-body problems, and allows us to clarify the dependence of the tunneling time on the system's parameters.  We also discuss the prospects of experimentally observing tunneling in the self-trapped regime.

The Bose-Hubbard dimer is mathematically equivalent to a spin system and to a certain limit of the Lipkin-Meshkov-Glick model.  Semiclassical quantization of these equivalent models was considered by~\cite{VanHemmen1986, VanHemmen1986a, Scharf1987} and~\cite{Enz1986}, respectively.  We complement these earlier works by providing a connection to the quantization condition of~\cite{Graefe2007}, proposing closed-form approximations valid in the relevant parameter range and offering a discussion of cold-atom experiments that could probe the tunneling phenomenon.

The rest of the paper is organized as follows.  In Section~\ref{sec:BH_dimer}, we review the Bose-Hubbard dimer and its mean-field approximation.  Section~\ref{sec:tunneling} is devoted to tunneling between the fixed points using exact diagonalization results.  In Section~\ref{sec:semiclassical_quantization}, we introduce the semiclassical quantization condition and obtain a closed form expression for the tunneling frequency.  Finally, in Section~\ref{sec:discussion} we discuss applications of this expression to problems of entanglement and atom loss rate from a dissipative optical lattice, as well as prospect of experimental confirmation.

\section{The Bose-Hubbard dimer}
\label{sec:BH_dimer}

We will consider bosonic atoms in a double well optical trap sufficiently deep that only the lowest state in each well is populated.  In this so-called two mode approximation, the atoms' dynamics is described by the Bose-Hubbard Hamiltonian~\cite{Milburn1997},
\begin{equation}
\hat{H} = -J(\hat{a}_1^\dagger \hat{a}_2 + \hat{a}_2^\dagger \hat{a}_1) + \frac{U}{2}\left(\hat{n}_1(\hat{n}_1 - 1) + \hat{n}_2(\hat{n}_2 - 1)\right)
\end{equation}
where $\hat{a}_i$ is the annihilation operator for a boson in well $i$ and $\hat{n}_i \equiv \hat{a}^\dagger_i \hat{a}_i$ is the number operator. 

Of special interest are the coherent states of the model~\cite{Arecchi1972, Hennig2012}.  These states correspond to all atoms being a single BEC~\cite{Trimborn2008, Trimborn2009} and can be characterized by their expectation values of the population imbalance between the wells, $z = (N_1-N_2)/2$, and of their relative phase, $\phi$.  In terms of the creation operators, the coherent states can be expressed as,
\begin{equation}
\ket{z,\, \phi} = \frac{1}{\sqrt{N}} \left(\sqrt{(1+z)/2}\,\hat{a}_1^\dagger + \sqrt{(1-z)/2}\,\mathrm{e}^{\imath\phi}\,\hat{a}_2^\dagger\right)^N \ket{0}.
\end{equation}
For large numbers of particles, the coarse dynamics of these states is well approximated by a bosonic Josephson junction (BJJ) model in which $z$ and $\phi$ are the dynamical variables.  The Hamiltonian of this model is,
\begin{equation}
\label{eq:mean-field}
\mathcal{H} = \frac{\Lambda z^2}{2} - \sqrt{1- z^2}\cos \phi,
\end{equation}
where $\Lambda\equiv \frac{UN}{2J}$ and the dimensionless time is $\tau = 2Jt/\hbar$~\cite{Raghavan1999}.  The BJJ model exhibits a bifurcation~\footnote{The general bosonic Josephson junction undergoes a hamiltonian saddle-node bifurcation~\cite{Howard2013} when the nonlinearity is increased. At the bifurcation, a new elliptically stable fixed point (a center or node) and a hyperbolically unstable fixed point (a saddle) come into being. In the special case of a symmetric double-well the location of the bifurcation in phase space coincides with the location of an already existing elliptically stable fixed point such that the bifurcation has the shape of a pitchfork as noted in~\cite{Zibold2010}.} at $\Lambda = 1$: as $\Lambda$ is increased beyond this critical value, a stable center at $z = 0$, $\phi = \pi$ breaks down into a saddle point point at the same coordinates and a pair of stable centers at $z = \pm \sqrt{1 - \frac{1}{\Lambda^2}}$, $\phi = \pi$.  These stable centers, corresponding to a persistent population imbalance between the dimer's two wells, are known as the self-trapping points.

\section{Tunneling between the self-trapping points}
\label{sec:tunneling}

\begin{figure}
\begin{center}
\includegraphics[width=\columnwidth]{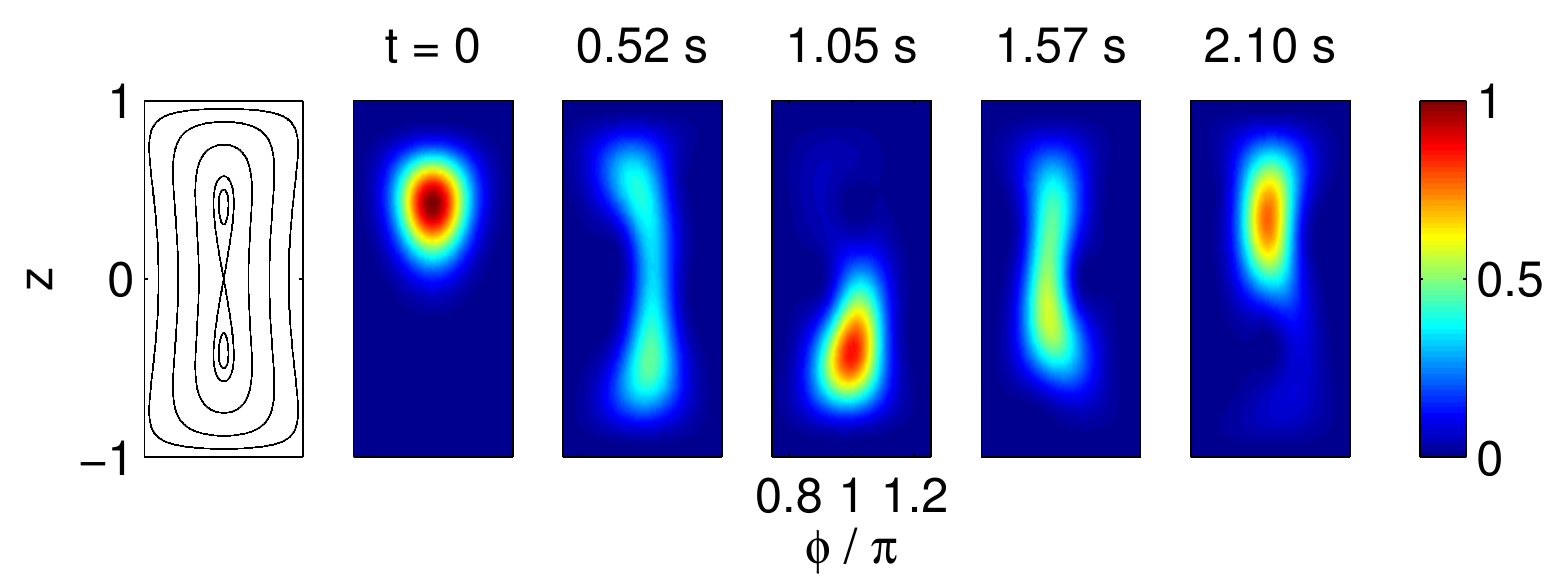}
\caption{Tunneling between the self-trapping fixed points.  In the BJJ model, trajectories sufficiently close to the self-trapping point remain confined to its neighborhood forever (far left panel).  However, as shown in the remaining panels, in the Bose-Hubbard model the Husimi function of a coherent state initially centered at the $z>0$ self-trapping fixed point tunnels from one fixed point to the other.  For the full video from which these stills are taken, see~\href{http://youtu.be/hX4nhoMb4G0}{http://youtu.be/hX4nhoMb4G0}. Parameters: $N = 40$ atoms, with $\Lambda = 1.1$ and $J = 10\unit{Hz}$.
\label{fig:husimi1.1}}
\end{center}
\end{figure}

Within the BJJ model, the self-trapping fixed points are stable: a trajectory initially sufficiently close to one of them remains close to it for all time.  In the full Bose-Hubbard dynamics, however, tunneling between the two self-trapping points occurs with a finite frequency.  An example of this process is shown in Figure~\ref{fig:husimi1.1}, which depicts the Husimi function~\cite{Lee1995}, a quasiprobability distribution over the coherent states $\ket{z,\,\phi}$ given by,
\begin{equation}
Q_\psi(z,\,\phi) = \left| \langle z,\,\phi | \psi \rangle \right|^2
\end{equation}
for a pure state $|\psi \rangle$.  The Husimi function is initially centered at one of the fixed points, but over time it tunnels to the other, and then back again.

A quantitative signature of the tunneling is an oscillation of the wells' populations.  The frequency of this oscillation can be found by numerically integrating the Schr\"{o}dinger equation of the Bose-Hubbard dimer for a long time and computing the power spectrum of the well populations.  The most prominent feature in the spectrum corresponds to the tunneling frequency.

\begin{figure}
\begin{center}
\includegraphics[width=\columnwidth]{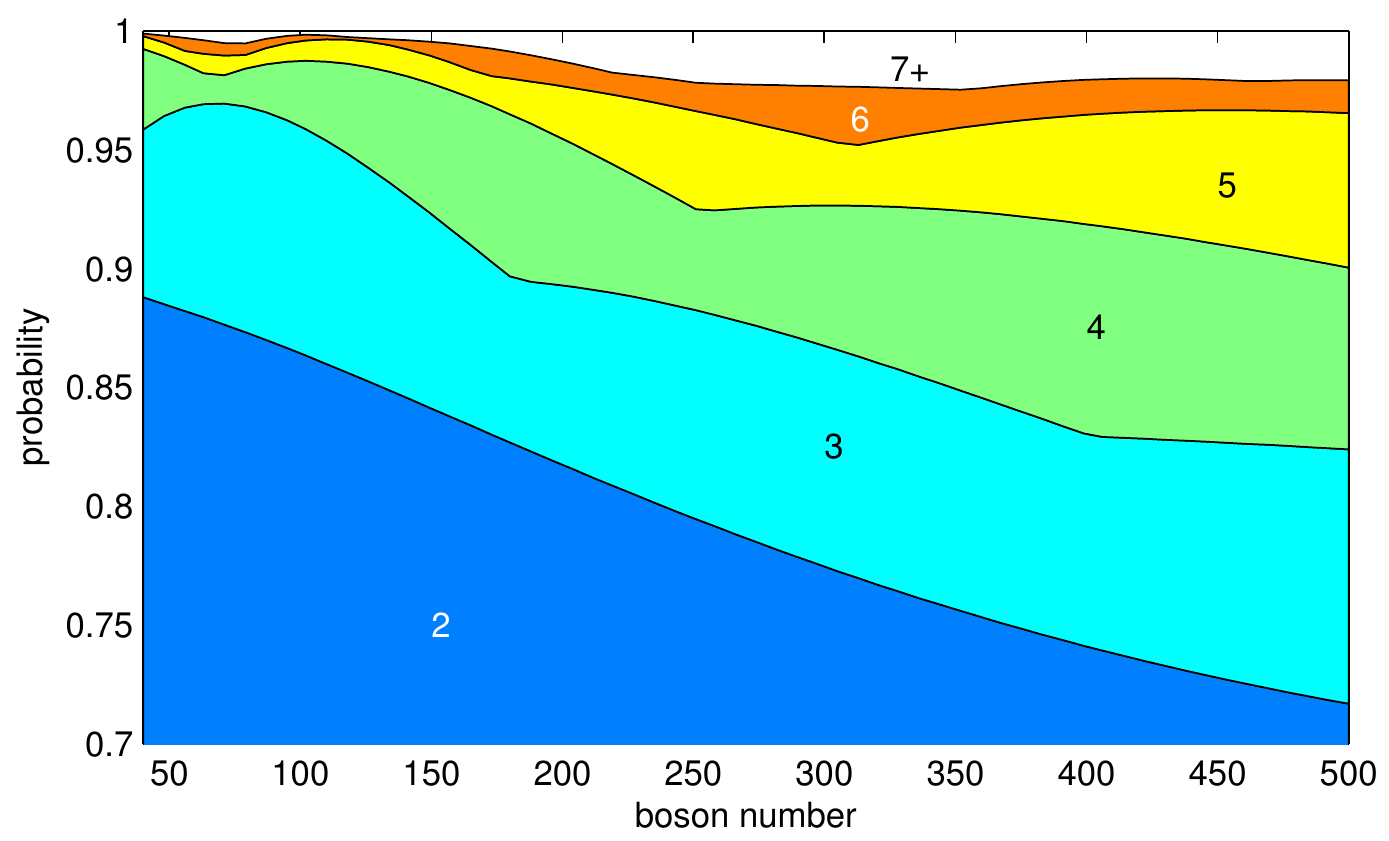}
\caption{The probability of observing the coherent state centered at the self-trapping fixed point in one of the $n$ most probable states, for $n = 2,\,3,\,4,\ldots$, as a function of the particle number $N$.  ($\Lambda = 1.025$, $U = 2\pi\times 0.063\unit{Hz}$.)
\label{fig:top_weights}}
\end{center}
\end{figure}

\begin{figure}
\begin{center}
\includegraphics[width=\columnwidth]{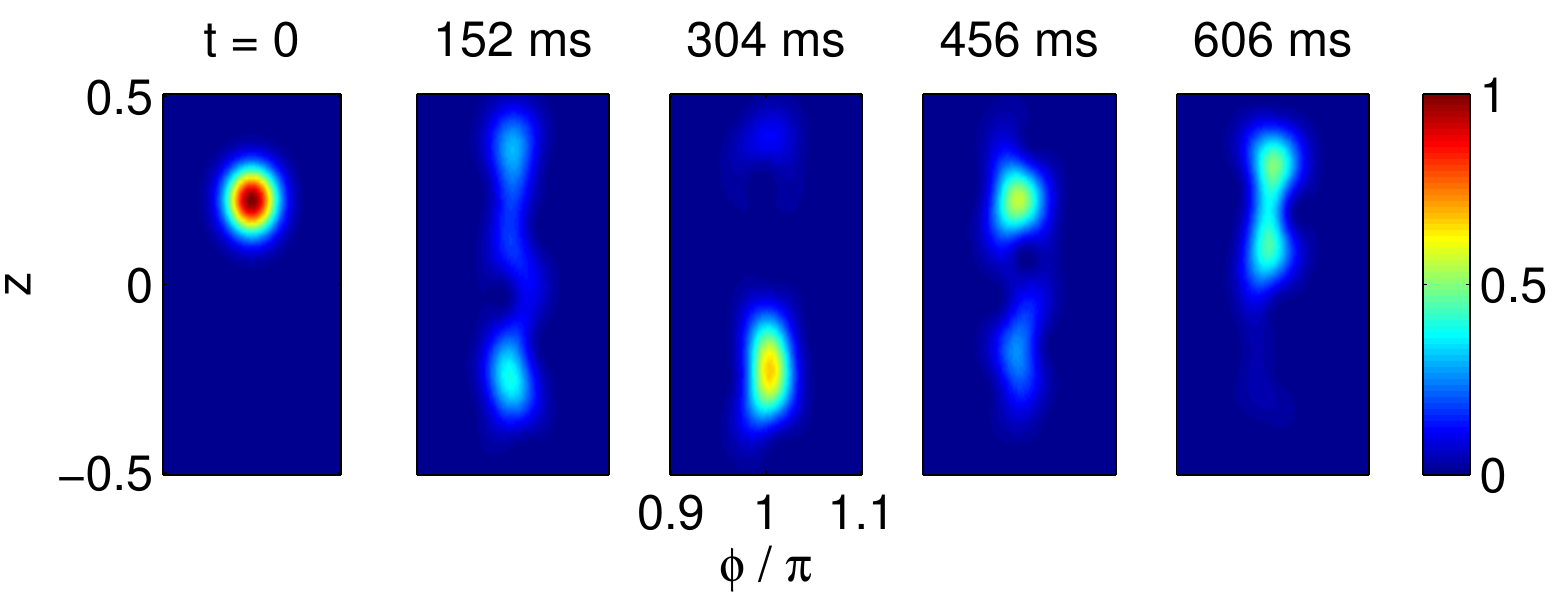}
\caption{A two-state description of the tunneling remains valid as $N$ increases, although the dynamics is more complex as the system becomes less discrete.  The Husimi function  is shown at the five times spaced by a quarter of the tunneling period expected from the two-state model.  For the full video from which these stills are taken, see~\href{http://youtu.be/p_LL85VBohU}{http://youtu.be/p\_LL85VBohU}.  ($N = 500$ atoms, $\Lambda = 1.025$ and $U = 2\pi \times 0.063\unit{Hz}$.)
\label{fig:husimi_high_N}}
\end{center}
\end{figure}

Since the dynamics of the coherent state near the self-trapping fixed points appears very simple, we may try to reduce the dimensionality of the problem by restricting the system to some subspace of the Hilbert space.  Remarkably, in the neighborhood of the mean-field fixed points, only a few energy eigenstates contribute appreciably to the coherent state~\cite{Chuchem2010, Pudlik2013}.  How many states need to be accounted for depends on the particle number (see Figure~\ref{fig:top_weights}).  Our intuition is that as $N$ increases, the ``size'' of the coherent state in phase space shrinks, but the ``size'' of the eigenstates shrinks even faster, and ever-more eigenstates are needed to correctly account for the coherent state dynamics.  However, even for a few hundred atoms much of the tunneling dynamics can be captured by keeping just two states (see Figure~\ref{fig:husimi_high_N}).  At the self-trapping fixed points, these two states are the pair of highest energy states of the Bose-Hubbard model~\footnote{The highest energy states are the relevant ones only if $\Lambda$ is positive.  Self-trapped fixed points also exist for $\Lambda \leq -1$, but in this case the relevant states are the two \emph{lowest} energy ones.}.  They are symmetric and antisymmetric combinations of states localized in each well.

\begin{figure}
\begin{center}
\includegraphics[width=0.8\columnwidth]{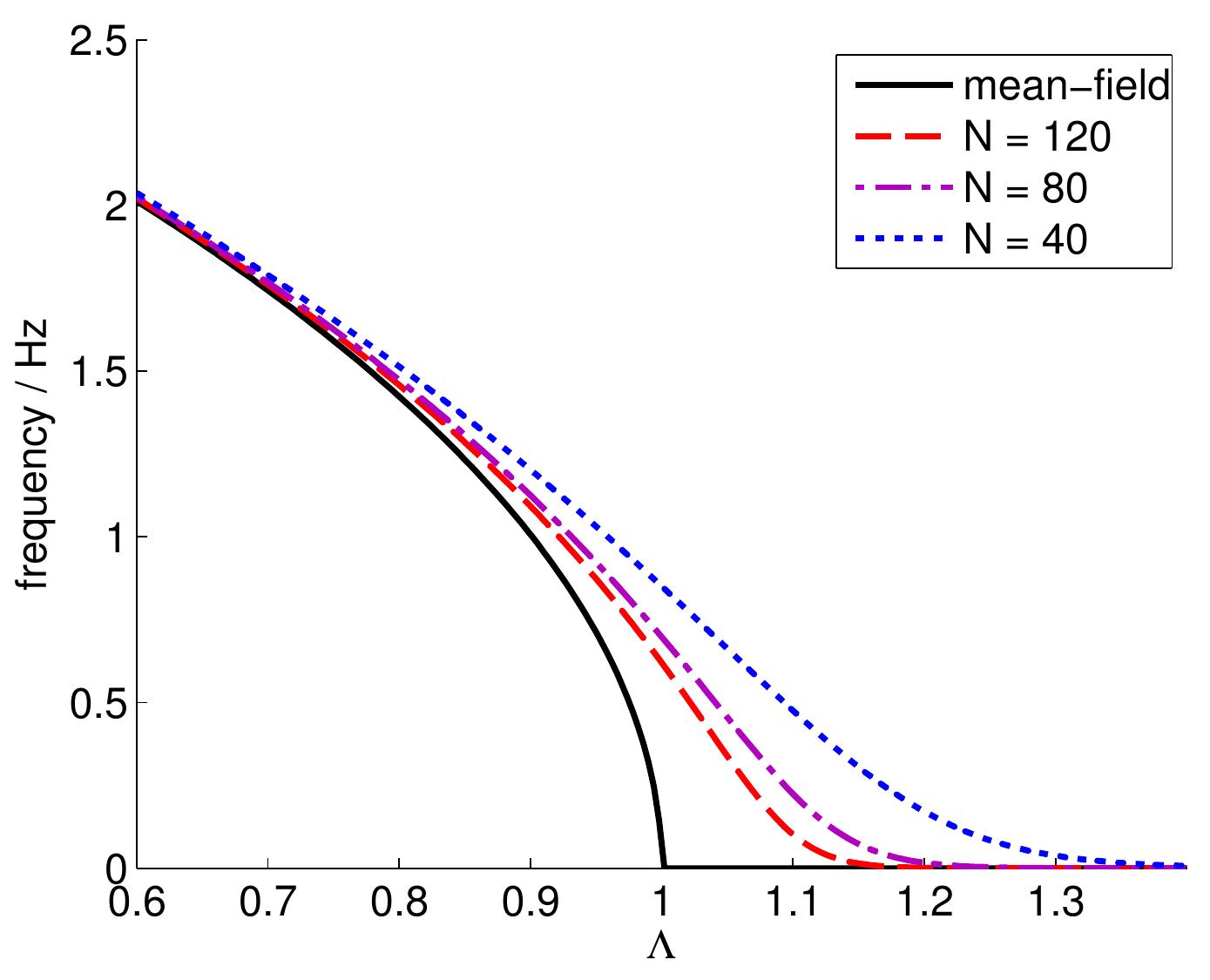}
\caption{The BJJ result of zero tunneling frequency for $\Lambda > 1$ is gradually approached by the two-eigenstate model as the number of atoms increases.  Nonetheless, a nonzero frequency is expected for any $\Lambda$ and any finite $N$.  In all plots, $J = 10\unit{Hz}$.
\label{fig:MF}}
\end{center}
\end{figure}

The energy splitting between the symmetric and antisymmetric states agrees closely with the oscillation frequency extracted by numerically integrating the Schr\"odinger equation~\footnote{See Supplementary Materials for a figure comparing the two-state prediction to numerical integration results.}.  The splitting between these states can also be computed for $\Lambda < 1$; in this case, there is only one fixed point at $\phi = \pi$, and the energy splitting closely agrees with the BJJ frequency of oscillations about that point.  Both above and below $\Lambda = 1$, the BJJ limit is approached as $N$ is increased (see Figure~\ref{fig:MF}).

The energies of the two highest-energy states are easily found numerically even for very large $N$, but it is desirable to explain the simple trends with $N$ and $\Lambda$ shown in Figure~\ref{fig:MF} using an analytical model.  Quantum perturbation theory can be used to obtain estimates of the tunneling frequency for small $J/U \approx N/\Lambda$~\cite{Bernstein1990,Salgueiro2007, Dounas-Frazer2007, Pudlik2013}, but not in the region $\Lambda \approx 1$ where tunneling becomes a significant effect.  In the next section, we will pursue an alternative approach.

\section{Semiclassical quantization} 
\label{sec:semiclassical_quantization}

To shed light on the convergence of the results of the two-state model to those of the BJJ, we will start with the BJJ model and recover additional features of the dynamics through Bohr-Sommerfeld quantization.  Graefe and Korsch~\cite{Graefe2007} applied Bohr-Sommerfeld quantization to this problem numerically, obtaining excellent estimates of the eigenenergies even for  atom numbers $N < 10$.  In this section, we start from their formulation of the quantization condition but proceed analytically to produce accurate closed-form expressions for the tunneling frequency.

The quantization condition in the self-trapping region of the symmetric dimer described by the Hamiltonian of Eq.~\refeq{eq:mean-field} is~\cite{Graefe2007},
\begin{equation}
\label{eq:GraefeQuantization}
\sqrt{1+\kappa^2} \cos (2S_w - S_\phi) = -\kappa.
\end{equation}
Here, $2S_w$ is the action associated with the self-trapped classical orbit, $\kappa = \exp(-\pi S_\epsilon)$ and $2S_\epsilon$ is the (Euclidean) action associated with tunneling.  Both $S_w$ and $S_\epsilon$ are measured in units of Planck's constant, $h$, and so are dimensionless~\footnote{The factors of 2 are conventional: the WKB approximation, which inspired this quantization condition, is typically expressed in terms of integrals $\int p\,dx$ between the turning points.  But $\int p\,dx = \frac{1}{2}\oint p\,dx = S/2$.}.  The phase correction term $S_\phi$ can be expressed in terms of $S_\epsilon$ as~\cite{Graefe2007},
\begin{equation}
\label{eq:df:S_phi}
S_\phi = \mathrm{arg} \,\Gamma\left(\frac{1}{2} + \imath S_\epsilon\right) - S_\epsilon \ln |S_\epsilon | + S_\epsilon.
\end{equation}
For a discussion of the physical significance of $S_\phi$, see~\cite[pp.~50--51]{Child1991}.

The actions $S_w$ and $S_\epsilon$ are functions of the energy $E$ and the nonlinearity $\Lambda$, and can be expressed as integrals over phase space (see Figure~\ref{fig:geometry_of_semiclassical_quantization}); this is discussed in greater detail in Appendix~\ref{sec:appendix_action_integrals}.
\begin{figure}
\begin{center}
\includegraphics[width=0.7\columnwidth]{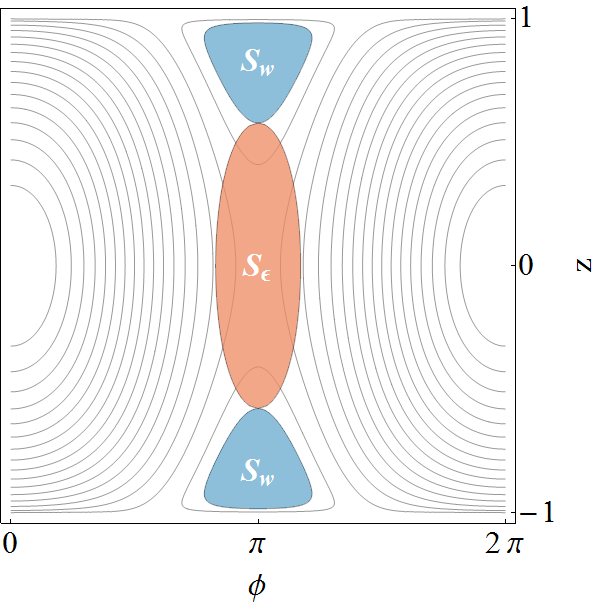}
\caption{The actions appearing in the quantization condition [Eq.~\refeq{eq:GraefeQuantization}] have a geometric interpretation.  This figure depicts the phase space of the BJJ model for $\Lambda = 2$.  The grey curves are trajectories; the actions $S_w$ and $S_\epsilon$ for energy $E = -1.15$ are equal to the areas of the marked regions.  In the case of $S_w$, the action corresponds to the phase space area of the classical orbit.
\label{fig:geometry_of_semiclassical_quantization}}
\end{center}
\end{figure}

Let us assume that the energy splitting between symmetric and antisymmetric combinations of states localized in the two self-trapping regions of phase space is small relative to the spacing of allowed energies in each region.  As shown in Appendix~\ref{sec:appendix_splitting_derivation}, in this case the quantization condition implies the splitting is approximately 
\begin{equation}
\label{eq:sc_splitting}
\Delta E = \frac{\hbar \omega}{\pi} \exp (\pi S_\epsilon),
\end{equation}
where $\omega$ is the frequency of the classical motion in a self-trapped orbit (related to the action of the orbit $S_w$, since $2\pi/\hbar\omega = T/\hbar = 2\partial S_w/\partial E$) and $S_\epsilon$ is as before the Euclidean action associated with the tunneling.  These quantities depend on the shape and size of the classical orbits, which are determined by $\Lambda$ and the energy of the unperturbed state $E$.

\begin{figure}
\begin{center}
\includegraphics[width=\columnwidth]{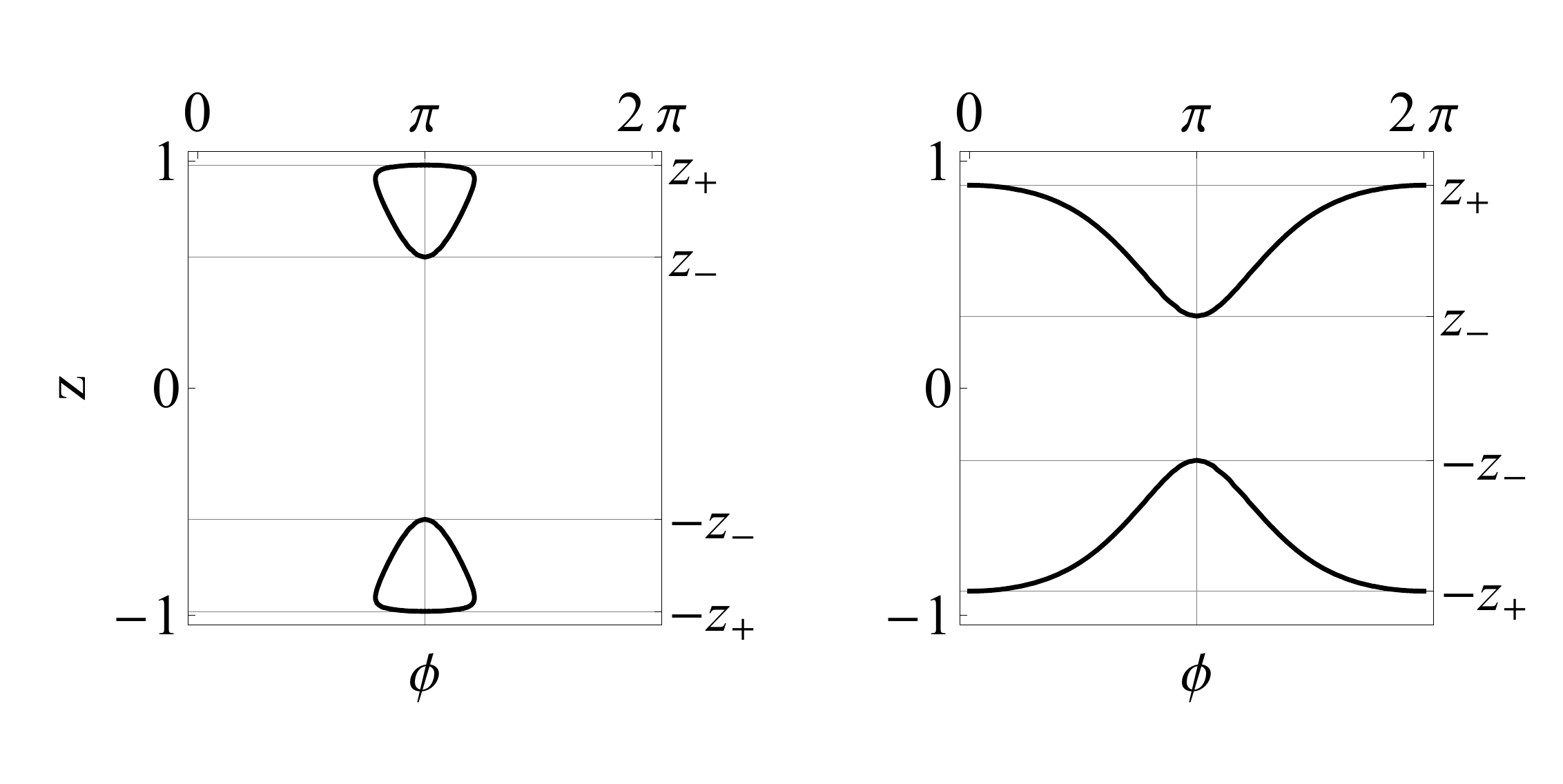}
\caption{Pairs of classical orbits and their turning points $z_\pm$.  The orbits on the left ($\Lambda = 2$, $E = 1.15$) are librations, while those on the right ($\Lambda = 4$, $E = 1.15$) are rotations.
\label{fig:turning_points}}
\end{center}
\end{figure}

Let the classical turning points be $z_\pm$ (see Figure~\ref{fig:turning_points}).  The size of the orbits is captured by the dimensionless parameter,
\begin{equation}
\label{eq:df:k}
k \equiv \sqrt{\frac{z_+^2 - z_-^2}{z_+^2}}.
\end{equation}
Furthermore, let,
\begin{equation*}
k' \equiv \sqrt{1-k^2} = \frac{z_-}{z_+},\quad\text{and}\quad
\alpha^2 = \frac{z_+^2 - z_-^2}{z_+^2 - 1}.
\end{equation*}
In Appendix~\ref{sec:appendix_action_integrals} we show that in terms of these quantities the splitting $\Delta E$ of the highest-energy state is given by,
\begin{widetext}
\begin{equation}
\label{eq:splittingFormula3}
\begin{split}
\Delta E &= \frac{\hbar \omega}{\pi}\exp (\pi S_\epsilon) \\
         &= \frac{\hbar z_+ \Lambda}{2\K(k)} \exp \left(-(N+1)\left(-\left(1- \frac{2E}{\Lambda}\right) \frac{1}{z_+} \Pi(z_+^{-2}, k') + z_+ \left(\K(k') - \E(k')\right)\right)\right),
\end{split}
\end{equation}
where $\K$, $\Pi$ and $\E$ are the complete elliptic integrals~\cite[\S 19.2(ii)]{DLMF}, while $E$ is the unperturbed highest-energy state energy satisfying the quantization condition,
\begin{equation}
\label{eq:quantization_condition}
\frac{\pi}{N+1} - \pi (1-z_+) \cdot\indicator{E < \Lambda/2}= \left(1 - \frac{2E}{\Lambda}\right)\frac{1}{z_+}\left(\K(k) - \frac{1}{1-z_+^2}\Pi(\alpha^2, k)\right) - z_+ \E(k),
\end{equation}
\end{widetext}
with $\indicator{\cdot}$ denoting the indicator function.

These complicated expressions constitute a solution to the problem of semiclassical quantization but offer little insight into the dimer's behavior.  Nonetheless, some of the problem's structure has become apparent:
\begin{enumerate}
\item The splitting depends on $E$ and $\Lambda$ only through the turning points $z_\pm$ and the combination $(1 - 2E/\Lambda)$.  The sign of this last quantity distinguishes between the two types of motion depicted in Figure~\ref{fig:turning_points}: $1 - 2E/\Lambda >0 $ for rotations (orbits surrounding one of the poles at $z = \pm 1$) and $1 - 2E/\Lambda < 0$ for librations.
\item The only nonelementary functions in the expressions above are the complete elliptic integrals $\K$, $\E$, and $\Pi$.  When they do appear they all take the same argument (modulus), either $k$ or $k'$, which is a measure of the size of the classical orbit.
\end{enumerate}
This structure can be exploited to find much simpler expressions for the splitting, valid in the limit of $N \gg 1$.

Let us first rescale the energy through a linear transformation:
\begin{equation}
\label{eq:df:e}
e = \left(-E + \frac{\Lambda}{2} + \frac{1}{2\Lambda}\right)\cdot \frac{(\Lambda-1)^2}{2\Lambda},
\end{equation}
The rescaled energy $e$ lies in $[0,\,1)$ for any orbit in the self-trapping region.  The highest-energy state orbit has an area $h/2$, while the total semiclassical action of a dimer with $N$ particles is $h(N+1)$.  As $N$ increases, both the highest-energy state energy $e$ and the dimensionless measure of orbit size $k$ [Eq.~\refeq{eq:df:k}] become small.  If the highest-energy state orbit is a libration ($e < (\Lambda-1)^{-2}$), expanding Eq.~\refeq{eq:quantization_condition} to lowest order in $k$ and $e$ and solving for $e$ gives an estimate of the highest-energy state energy,
\begin{equation}
\label{eq:approx_ground_state}
e \approx \frac{2\Lambda \sqrt{\Lambda^2 - 1}}{(\Lambda - 1)^2 (N+1)}.
\end{equation}
This estimate is very good: the relative error in approximating the numerical semiclassical result is less than 1\% for $N = 20$ and $\Lambda = 1.25$, and decreases with both $N$ and $\Lambda$.  Analogous expansions for the classical orbital frequency and the tunneling phase lead to the following expression for the ground state splitting:
\begin{equation}
\label{eq:approximate_splitting}
\Delta E \approx 2J\frac{\omega}{\pi}\left(\frac{1}{\omega}\mathrm{e}^{-z_0}\right)^{(N+1)(1-e)},
\end{equation}
where $z_0 \equiv \sqrt{1 - \frac{1}{\Lambda^2}}$ is the position of the self-trapping fixed point and $\omega = \sqrt{\Lambda^2 - 1}$ is the frequency of motion about it.  The tunneling frequency $\Delta E/\hbar$ decreases exponentially with the ``barrier width'' $\approx z_0$, the ``barrier height'' $\approx (1-e)$ and the number of atoms $N$.  The details of the calculation are described in Appendix~\ref{sec:appendix_approximate_quantization}.

\begin{figure}
\includegraphics[scale=0.5]{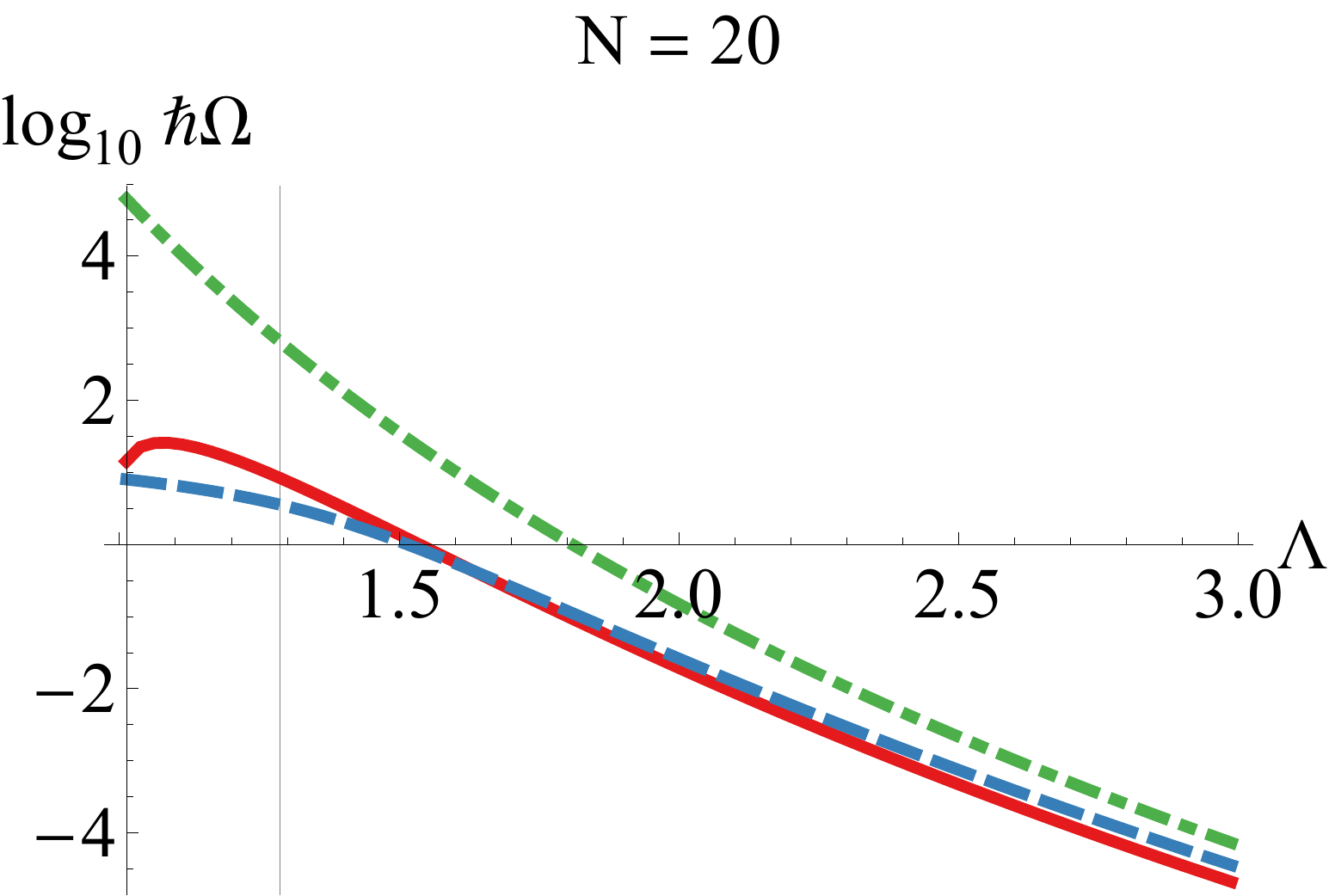}
\includegraphics[scale=0.5]{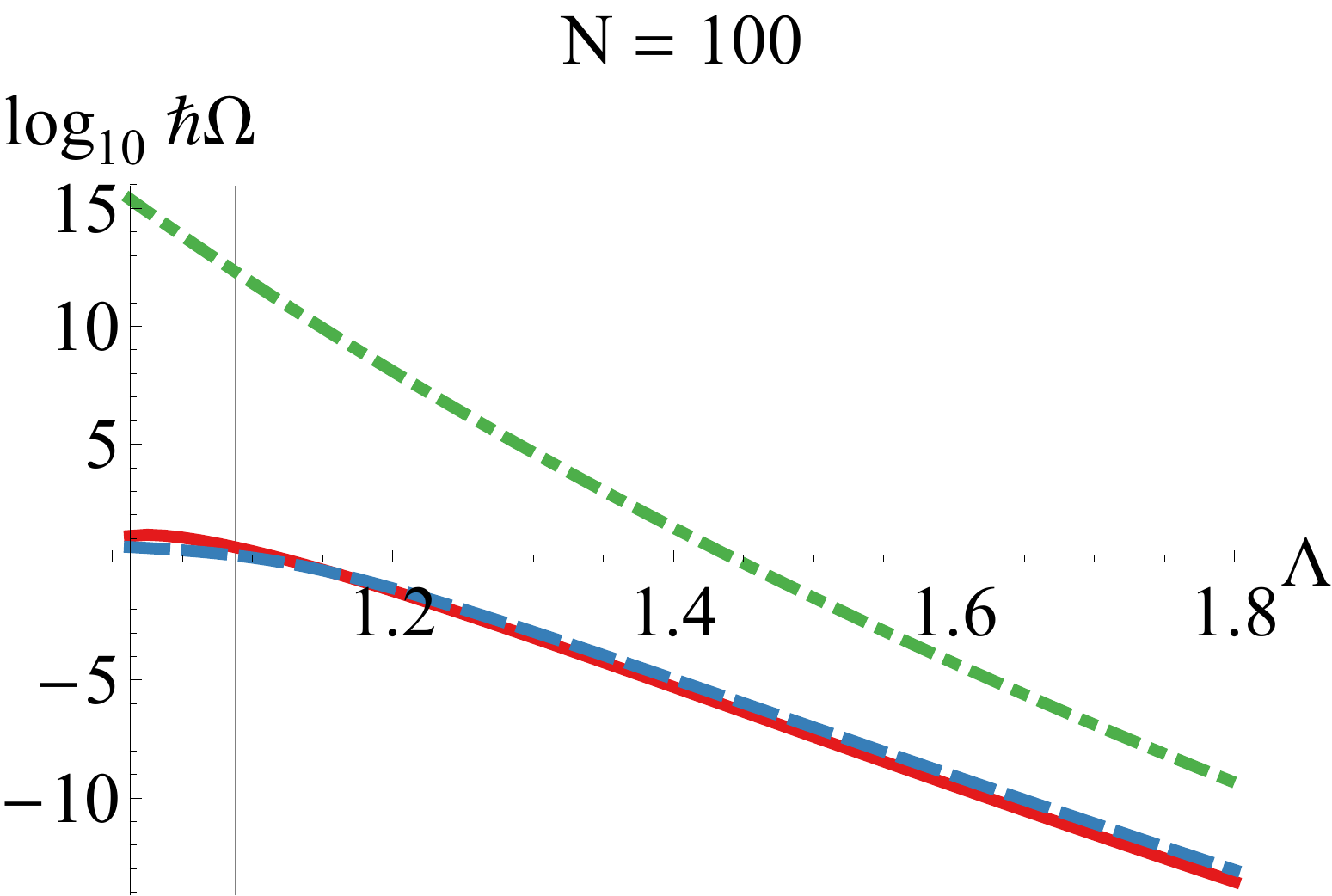}
\caption{Comparison of semiclassical estimates of the splitting with exact diagonalization.  The analytical approximation of Eq.~\refeq{eq:approximate_splitting} (red line) agrees closely with the results of exact diagonalization (blue dashed line).  In contrast, the approximation of~\cite{Scharf1987} (green dot-dashed line) performs poorly in this low-$\Lambda$ regime, especially for larger $N$.   The black vertical line marks the $\Lambda$ value below which the semiclassical approximation breaks down because the area of phase space associated with the self-trapped region is less than $h/2$.
\label{fig:approx_splitting_BH}}
\end{figure}

Figure~\ref{fig:approx_splitting_BH} compares the semiclassical splitting estimates with the results of exact diagonalization of the Bose-Hubbard model.  The results of solving the quantization problem numerically are not shown: except for $\Lambda$ so small that not even one semiclassical orbit fits within the self-trapping region, they agree very closely with the exact Bose-Hubbard splitting.  The analytic approximation discussed in this section is generally within a factor of 2 of the exact result, and improves with $N$.  Since the splitting changes by as many as 15 orders of magnitude over the investigated range of $\Lambda$, this agreement amounts to remarkably robust performance.

A different closed-form semiclassical approximation to $\Delta E$ was obtained in~\cite{VanHemmen1986} and refined in~\cite{Scharf1987}.  This last approximation attains an excellent accuracy, on the order of a few percent, but only for $U \approx J$.  In the context of cold atomic experiments, in which the atom number is on the order of hundreds, this corresponds to astronomically small tunneling frequencies (well below $10^{-100}$ Hz).  For $U \ll J$, where the tunneling frequency becomes large, the approximation of~\cite{Scharf1987} is many orders of magnitude from the true value (see Figure~\ref{fig:approx_splitting_BH}).  Therefore, the approximation we provide in Eq.~\refeq{eq:approximate_splitting} is the first closed-form expression valid in the experimentally relevant regime.

\section{Discussion}
\label{sec:discussion}


In this section, we consider the implications of the analysis presented above for three problems: determining the time scale for macroscopic entanglement, producing quantum speedup of dissipation, and obtaining experimental confirmation.

\subsection{Time scale for macroscopic entanglement}

Tunneling in the self-trapping regime leads to the generation of entangled superpositions of many-particle states, or macroscopic entanglement~\cite{Carr2010}.  The entanglement between the two modes is maximized at times $T/4$ and $3T/4$, where $T$ is the tunneling period.  Therefore, our semiclassical estimate of the tunneling frequency immediately yields an estimate of the time required for entanglement generation.  It is notable that the dynamics of entanglement, a profoundly unclassical phenomenon, is captured by the first quantum correction to the (classical) BJJ model.

\subsection{Quantum speedup of dissipation}

So far we have considered only an isolated Bose-Hubbard dimer.  We will now discuss the central role tunneling in the self-trapped regime plays in a quantitatively accurate model of a dissipative dimer that leaks atoms to the environment.

Consider a coherent state of $N$ bosons centered at one of the self-trapping fixed points, say the left well.  We will attempt to model its dynamics within a two-dimensional subspace of the full system's Hilbert space, the subspace spanned by the symmetric and antisymmetric energy eigenstates, $\ket{E_S}$ and $\ket{E_A}$.  In the basis of states localized in the two wells, $\ket{1} = (\ket{E_S} + \ket{E_A})/\sqrt{2}$ and $\ket{2} = (\ket{E_S} - \ket{E_A})/\sqrt{2}$, the  Hamiltonian is represented by the matrix,
\begin{equation*}
\begin{pmatrix}
\bar{E} & \Delta E \\
\Delta E  & \bar{E}
\end{pmatrix}
\end{equation*}
where $\bar{E} = (E_S + E_A)/2$ and $\Delta E = (E_S - E_A)/2$.  These parameters can be calculated semiclassically with high accuracy as we have shown in the preceding section [Eq.~\refeq{eq:approx_ground_state} and Eq.~\refeq{eq:approximate_splitting}], though we use exact values in the simulation discussed below.  The initial condition is the localized state $\ket{1}$.  Now, assume there is decay from the right well at a rate $\gamma$.  In the two-level model this is described by the effective decay rates,
\begin{equation*}
\Gamma_1 = -\gamma \melement{1}{\hat{a}_2^\dagger \hat{a}_2}{1}, \quad \Gamma_2 = -\gamma \melement{2}{\hat{a}_2^\dagger \hat{a}_2}{2},
\end{equation*}
leading to the effective Hamiltonian,
\begin{equation}
\label{eq:dissipative_hamiltonian_eff}
H_\mathrm{eff}^{(2)} = \begin{pmatrix}
\bar{E} - \imath\Gamma_1/2 & \Delta E \\
\Delta E                   & \bar{E} - \imath\Gamma_2/2
\end{pmatrix}.
\end{equation}
This simple model can be used to estimate how the probability of all $N$ atoms remaining in the system diminishes over time.  To evaluate the results, we compare them to those obtained using the complete coherent state and the full master equation~\cite{Witthaut2011, Witthaut2008, Kordas2012},
\begin{equation}
\label{eq:dissipative_hamiltonian_bh}
\dot{\hat{\rho}} = -\imath [\hat{H}, \hat{\rho}] - \frac{\gamma}{2}\left(\hat{a}_1^\dagger \hat{a}_1 \hat{\rho} + \hat{\rho} \hat{a}_1^\dagger \hat{a}_1 - 2 \hat{a}_1 \hat{\rho} \hat{a}_1^\dagger \right).
\end{equation}

\begin{figure}
\includegraphics[width=0.8\columnwidth]{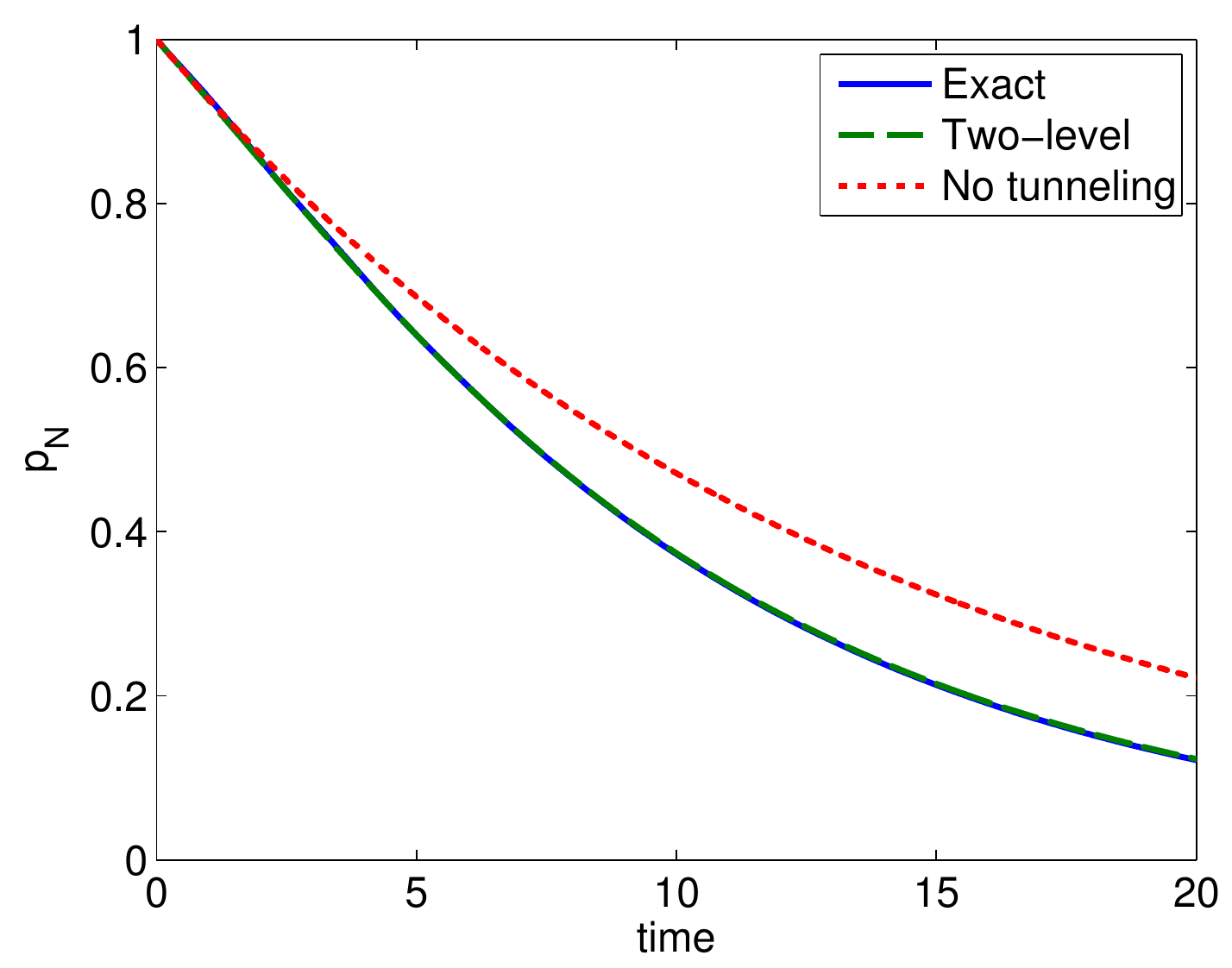}
\caption{Correctly estimating the rate of tunneling between the self-trapping fixed points is critical to predicting the atom loss rate from a leaky dimer.  The probability of finding all $N$ atoms in the system over time is plotted for three different models.  The dashed green line is the simple Hamiltonian of Eq.~\refeq{eq:dissipative_hamiltonian_eff}, based only on two parameters $\bar{E}$ and $\Delta E$ which can be calculated semiclassically.  It overlaps with the numerically exact results obtained by integrating the many-body master equation of Eq.~\refeq{eq:dissipative_hamiltonian_bh} (solid blue line).  The simple model with $\Delta E$ set to zero differs significantly (dotted red line). ($J = 1\unit{Hz}$, $U = 4/5\unit{Hz}$, $N = 6$) \label{fig:dissipation}}
\end{figure}

The probabilities of remaining in the $N$ atom subspace predicted using the two Hamiltonians are shown in Figure~\ref{fig:dissipation}. 
 If many-body tunneling between the fixed points is neglected ($\Delta E = 0$), the rate of atom loss is significantly underestimated.  But when the correct value of $\Delta E$ is used, the effective two-state model produces results almost indistinguishable from the full Bose-Hubbard.  Remarkably, we can thus reproduce the decay dynamics of a correlated many-body system using only two parameters, $\bar{E}$ and $\Delta E$, which can be calculated semiclassically.

\subsection{Prospects of experimental observation}

The BJJ dynamics of the BEC dimer was experimentally mapped out in great detail a few years ago~\cite{Zibold2010}.  Could a similar experiment observe tunneling between the fixed points for $\Lambda > 1$?

Experimental realizations of the dimer fall into two categories: ``external'' and ``internal''~\cite{Leggett2001}, or those utilizing two spatially separated wells and those using two internal states of atoms.  Correctly describing the dynamics of the spatially separated wells requires going beyond the Bose-Hubbard model that was our starting point in this work, as the localized orbitals associated with the operators $\hat{a}_i$, $\hat{a}_i^\dagger$ are time-dependent~\cite{Sakmann2009}.  Fortunately, this complication does not arise in the case of internal states~\cite{Zibold2012}.  Therefore, the tunneling and dissipation enhancement effects we have described are most likely to be observed in experiments relying on internal states.

\begin{figure}
\includegraphics[width=0.8\columnwidth]{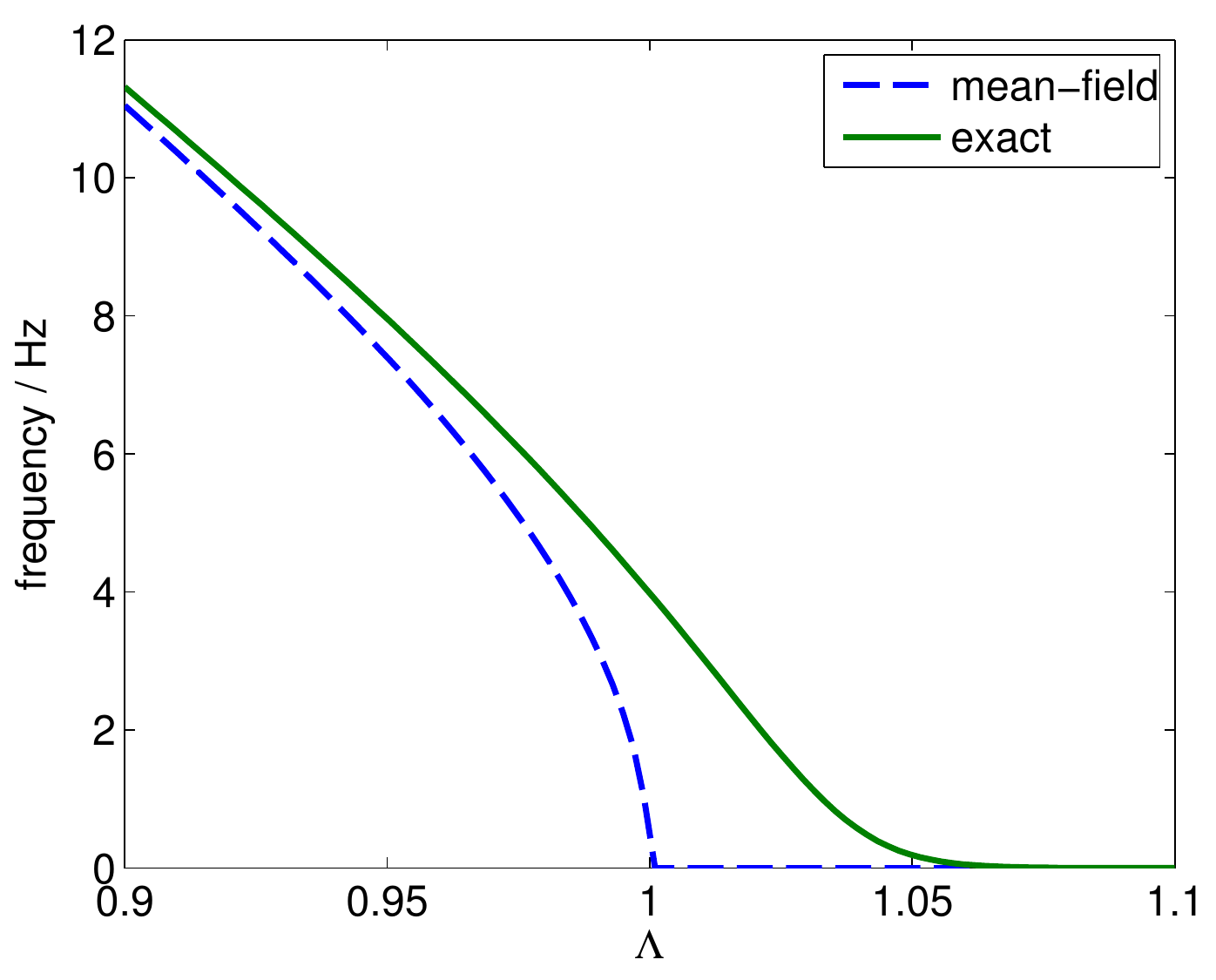}
\caption{Frequency of tunneling between the fixed points versus $\Lambda$ for $U = 2\pi\times 0.063~\unit{Hz}$ and $N = 500$, the experimental parameters of~\cite{Zibold2010}.  The mean-field prediction is also shown for reference.\label{fig:experimental_prediction}}
\end{figure}

The expected tunneling frequency given the experimental parameters of~\cite{Zibold2010} is shown in Figure~\ref{fig:experimental_prediction}.  The frequency is on the order of a few Hertz.  Since the atom decay times reported in this experiment are $\sim 100$ ms, the tunneling frequency is too small to be observed at present.  However, an order of magnitude improvement in atom retention times would render experimental observation feasible.

At first glance, it may seem that the retention time limitation could be sidestepped by lowering both $N$ and $J$ by the same factor.  Since the quantum tunneling time depends on $N$ exponentially, but on $J$ only linearly  [Eq.~\refeq{eq:approximate_splitting}], this could speed up the semiclassical dynamics while keeping $\Lambda$ constant.  Unfortunately, the experiment of~\cite{Zibold2010} was already carried out at the lowest $J$ currently accessible: lowering it even more introduces unacceptable noise due to EM fluctuations~\footnote{We thank Wolfgang Muessel for private communication on this point.}.

\subsection{Beyond the dimer: semiclassical quantization for lattices}

Although our analysis was limited to the dimer, analogous processes should occur in a system with multiple states, only one of which has an appreciable population.  The Bose-Hubbard Hamiltonian can be straightforwardly extended to such systems; in the case of the trimer, self-trapping has been demonstrated in both the quantum model and its classical limit~\cite{Mossmann2006,Hennig2010}.  However, carrying out semiclassical quantization is difficult because the classical model is now chaotic.  So far, progress has only been made for the case of very small and very large $J/U$~\cite{Itin2011}, i.e.~precisely the region of parameter space where tunneling between the self-trapping points does not take place.  Therefore, the extension of our results beyond the dimer is likely to prove challenging.

\section{Summary \& Outlook}
\label{sec:conclusion}

We have studied the tunneling between the self-trapped fixed points of the BEC dimer using a semiclassical approach.  We derived an exact solution to the problem in terms of elliptic integrals giving the phase space areas of semiclassical orbits.  For particle numbers $N \gg 1$, the semiclassical ground state orbit and (appropriately transformed) energy become small; in this limit we found an approximate closed-form expression for the tunneling frequency that is accurate in the experimentally relevant parameter range.  The tunneling frequency decreases exponentially with the effective width and height of barriers in phase space, as well as the number of particles.  Nonetheless, accounting for the tunneling is crucial to obtaining quantitatively accurate estimates of atom loss rates in a leaky dimer.

\begin{acknowledgments}
We wish to thank Wolfgang Muessel, Markus Oberthaler, Kaspar Sakmann, Andrea Trombettoni, Stephanos Venakides, and Tilman Zibold for helpful discussions.  We are also grateful for the hospitality of Joshua E.~S.~Socolar and the Duke University Physics Department.

This work was supported in part by Boston University.
\end{acknowledgments}

\appendix

\section{Derivation of Equation~\refeq{eq:sc_splitting}}
\label{sec:appendix_splitting_derivation}

In this appendix, we use Eq.~\refeq{eq:GraefeQuantization}, the quantization condition of Graefe and Korsch~\cite{Graefe2007}, to derive an approximate expression for the energy splitting of the nearly-degenerate self-trapped eigenstates.  This expression and its derivation have been known to scholars of the WKB approximation (see~\cite{Razavy2003}, p.~49, or~\cite{Child1991}, p.~52), but the discussion we give here is more complete than that found in other sources.

\begin{figure}
\begin{center}
\includegraphics[width=\columnwidth]{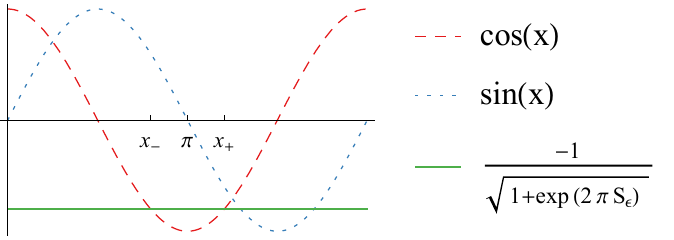}
\caption{Graphical representation of the roots of Eq.~\refeq{eq:GraefeQuantization2}.
\label{fig:xpm}}
\end{center}
\end{figure}

Eq.~\refeq{eq:GraefeQuantization} can be rewritten as,
\begin{equation}
\label{eq:GraefeQuantization2}
\cos (2S_w - S_\phi) = - \frac{1}{\sqrt{1 + \exp(2\pi S_\epsilon)}}.
\end{equation}
Considered as a function of $x \equiv 2S_w - S_\phi$, this equation has pairs of solutions symmetrically spaced about $(2n+1)\pi$ (see Figure~\ref{fig:xpm}).  The pairs of roots coalesce as $S_\epsilon \to -\infty$: in the absence of tunneling, states come in degenerate pairs, one localized in each well.  Let the two solutions near $x=\pi$ be $x_\pm$, with $x_+ > \pi$ and $x_- < \pi$. 
 We have,
\begin{equation}
\label{eq:GraefeQuantization3}
\tan x_\pm = \frac{\mp \sqrt{1-\cos^2(x_\pm)}}{\cos x_\pm} = \mp \exp(\pi S_\epsilon),
\end{equation}
where the sign difference on the right-hand-side arises because $\sin(x)$ changes sign at $x=\pi$, between $x_-$ and $x_+$.

Recall that $x \equiv 2S_w - S_\phi$ is a function of energy.  Assume the ground state energy splitting $\Delta E$ is sufficiently small that $x(E)$ is approximately linear in an interval of width $\Delta E$ about the ground state energy, $E_0$.  Then,
\begin{equation*}
x_\pm = x(E_0 \pm \Delta E/2),
\end{equation*}
and Eq.~\refeq{eq:GraefeQuantization3} gives,
\begin{multline*}
\tan (2S_w(E_0 \pm \Delta E/2) - S_\phi(S_\epsilon(E_0 \pm \Delta E/2))) = \\ \mp \exp(\pi S_\epsilon(E_0 \pm \Delta E/2)),
\end{multline*}
or,
\begin{multline*}
2S_w(E_0 \pm \Delta E/2) - S_\phi(S_\epsilon(E_0 \pm \Delta E/2)) = \\ \mp \arctan \left(\exp \pi S_\epsilon(E_0 \pm \Delta E/2)\right).
\end{multline*}
Expanding to first order about $E_0$,
\begin{multline*}
2S_w - S_\phi \pm \left(2 \frac{\partial S_w}{\partial E} - \frac{\partial S_\phi}{\partial S_\epsilon} \frac{\partial S_\epsilon}{\partial E}\right) \frac{\Delta E}{2} = \\ \mp \arctan(\exp (\pi S_\epsilon)) - \frac{2\pi}{\cosh (\pi S_\epsilon)} \frac{\partial S_\epsilon}{\partial E} \frac{\Delta E}{2}.
\end{multline*}
Subtracting the lower signs from the upper signs and rearranging yields,
\begin{equation}
\label{eq:splittingFormula1}
\frac{\Delta E}{2} = - \frac{\arctan \exp (\pi S_\epsilon)}{2\frac{\partial S_w}{\partial E} - \frac{\partial S_\phi}{\partial S_\epsilon}\frac{\partial S_\epsilon}{\partial E}}.
\end{equation}
Consider the second term in the denominator.  Letting $\xi \equiv S_\epsilon$ and using the definition of $S_\phi$ [Eq.~\refeq{eq:df:S_phi}], the unitless derivative can be written as,
\begin{equation}
\frac{\partial S_\phi}{\partial S_\epsilon} = - \ln \xi + \frac{1}{2} \psi\left(\frac{1}{2} - \imath \xi\right) + \frac{1}{2} \psi\left(\frac{1}{2} + \imath \xi\right),
\end{equation}
where $\psi$ is the digamma function, defined as
\begin{equation*}
\psi(t) = \frac{\Gamma'(t)}{\Gamma(t)}.
\end{equation*}
For $|t| > 3$, excellent approximation (good to 0.03\%) to this function is provided by the asymptotic expansion~\cite[5.11.2]{DLMF},
\begin{equation*}
\psi(t) \approx \ln t - \frac{1}{2t} - \frac{1}{12t^2}.
\end{equation*}
Using this expansion,
\begin{equation*}
\begin{split}
\frac{\partial S_\phi}{\partial S_\epsilon} &\approx \frac{1}{2} \ln \left(1 + \frac{1}{4\xi^2}\right) - \frac{4}{3} \frac{1+2\xi^2}{(1+4\xi^2)^2} \\
 &\approx \frac{3 - 8\xi^2(1-2\xi^2)}{24\xi^2(1+2\xi^2)^2}.
\end{split}
\end{equation*}
This expression is already smaller than 0.01 at $\xi = 2$, and decreases with $\xi$ as $1/\xi^2$.  Since the phase space derivatives $\partial S_w/\partial E$ and $\partial S_\epsilon/\partial E$ are of the same order, and $\xi = S_\epsilon$ is of order $N$, the second term in the denominator of Eq.~\refeq{eq:splittingFormula1} can be neglected:
\begin{equation*}
\Delta E = - \frac{\arctan \exp \pi S_\epsilon}{\frac{\partial S_w}{\partial E}}.
\end{equation*}

Since the splitting is small, $\exp (\pi S_\epsilon) \ll 1$ and so $\arctan \exp (\pi S_\epsilon) \approx \exp (\pi S_\epsilon)$.  If we let $T = 2\pi/\omega$ be the period of the orbit corresponding to the action $2S_w$,
\begin{equation*}
2\frac{\partial S_w}{\partial E} = \frac{1}{\hbar} T = \frac{2\pi}{\hbar \omega}.
\end{equation*}
Neglecting the second term in the denominator of Eq.~\refeq{eq:splittingFormula1}, we get,
\begin{equation}
\label{eq:splittingFormula2}
\Delta E = - \frac{\hbar\omega}{\pi} \exp (\pi S_\epsilon).
\end{equation}
The negative sign of $\Delta E$ indicates that $x_+$ is actually lower in energy than $x_-$.

As a special case, this result applies to a single particle in a double-well potential described by the Sch\"{o}dinger equation.  For that special case there exist a simpler derivation of Eq.~\refeq{eq:splittingFormula2}: see~\cite{Landau1981}, \S 50.

\section{Elliptic integral expressions for $T$, $S_w$ and $S_\epsilon$}
\label{sec:appendix_action_integrals}

To perform actual calculations using the formula,
\begin{equation*}
\Delta E = \frac{\hbar \omega}{\pi} \exp (\pi S_\epsilon),
\end{equation*}
we need to find explicit expressions for $\omega$ (or the corresponding period $T$) and $S_\epsilon$ in terms of $E$ and $\Lambda$.  It will also prove useful to find an expression for $2S_w$, the action associated with the self-trapped orbit, which determines the energy about which the splitting takes place.  All of these quantities depend on the shape of the classical orbits of the mean-field Hamiltonian of Eq.~\refeq{eq:mean-field}.  The equation of the orbit is,
\begin{equation}
\label{eq:orbit}
\phi(z,E,\Lambda) = \arccos \frac{\Lambda z^2-2E}{2\sqrt{1-z^2}},
\end{equation}
and the classical turning points of the orbits (see Figure~\ref{fig:turning_points}) are,
\begin{equation}
\label{eq:zpm}
z_\pm(E,\Lambda) = \sqrt{\frac{\pm \sqrt{1 - 2E\Lambda + \Lambda^2} + \Lambda E - 1}{\Lambda^2/2}}.
\end{equation}
In what follows, we will generally suppress the explicit dependence of $\phi$ and $z_\pm$ on $E$ and $\Lambda$ to obtain clearer expressions.  Recall that we defined the dimensionless measure of orbit size as,
\begin{equation*}
k \equiv \sqrt{\frac{z_+^2 - z_-^2}{z_+^2}}.
\end{equation*}
We begin with the simplest problem, that of deriving an expression for the orbit period $T$.  The approach to computing the action integrals $S_\epsilon$ and $S_w$ is the same, but the technical details are more involved.

See~\cite{Graefe2014} and the references therein for a deeper look at the geometry of the classical model and its relationship to Bose-Hubbard dynamics.

\subsection{Period $T$ of the classical orbit}

The equation of motion for $z$ is,
\begin{equation}
\dot{z} = -\frac{\partial H}{\partial \phi} = -\sqrt{1-z^2} \sin \phi,
\end{equation}
and so the period is,
\begin{equation}
T = 2 \left|\int_{z_-}^{z_+} \frac{dt}{dz}\,dz \right| = 2 \int_{z_-}^{z_+} \frac{dz}{\sqrt{1-z^2} \sin \phi(z)}.
\end{equation}
Since $\sin (\arccos x) = \sqrt{1-x^2}$, we can use Eq.~\refeq{eq:orbit} to eliminate the trigonometric functions:
\begin{equation}
T = 4 \int_{z_-}^{z_+} \frac{dz}{\sqrt{4(1-z^2) - (\Lambda z^2 - 2E)^2}}.
\end{equation}
Although at first glance this expression has a very complicated structure, the polynomial in the denominator (which is also encountered in the $S_w$ and $S_\epsilon$ integrals) can be rewritten in the more suggestive form,
\begin{equation}
\label{eq:elliptic_polynomial}
4(1-z^2) - (\Lambda z^2 - 2E)^2 = -\Lambda^2(z^2 - z_+^2)(z^2 - z_-^2).
\end{equation}
The period is therefore,
\begin{equation}
\begin{split}
T &= \frac{4}{\Lambda}\int_{z_-}^{z_+} \frac{dz}{\sqrt{(z_+^2 - z^2)(-z_-^2 + z^2)}}\\
 &= \frac{4}{\Lambda z_+} \K\left(\sqrt{\frac{z_+^2-z_-^2}{z_+^2}}\right) = \frac{4}{\Lambda z_+} \K(k),
\end{split}
\end{equation}
where $\K$ is the complete elliptic integral of the first kind.  Note that in this expression, time is measured in the dimensionless units introduced with the Hamiltonian of Eq.~\refeq{eq:mean-field}.  Converting the units to seconds,
\begin{equation}
T = \frac{1}{J}\frac{2}{\Lambda z_+} \K(k),
\end{equation}
where $J$ is measured in Hertz.

\subsection{Action of the classical orbit}

The phase space areas (and so actions) associated with the classical orbits can be found by integrating $\phi(z)$.  For an orbit in the self-trapping region, the action is
\begin{equation}
\label{eq:action}
\begin{split}
S(E,\Lambda) &= h\frac{N+1}{4\pi}\cdot \Bigg(2 \int_{z_-}^{z_+} \pi - \phi(z) \,dz \\
&\quad + 2\pi(1-z_+)\indicator{E < \Lambda/2}\Bigg).
\end{split}
\end{equation}
The prefactor $h\frac{N+1}{4\pi}$ normalizes the total area of phase space to be $h(N+1)$, with $N$ the number of particles.  If $E < \Lambda/2$, the orbit is a rotation orbit (see Figure~\ref{fig:turning_points}) and the area of the ``cap'' at $\abs{z} > z_+$ is added to the integral of $\phi(z)$.

The integral in Eq.~\refeq{eq:action} can be simplified through an integration by parts:
\begin{equation*}
\int_{z_-}^{z_+} \pi - \phi(z)\,dz = \int_{z_-}^{z_+} \frac{z^2 \left(z^2 + 2\frac{E-\Lambda}{\Lambda}\right)}{(1-z^2)\sqrt{(z^2-z_-^2)(-z^2+z_+^2)}},
\end{equation*}
where the boundary term is zero since  $\phi(z_\pm) = \pi/2$.  This is an elliptic integral~\cite[\S 19.2(i)]{DLMF} and can be reduced to the canonical elliptic integrals using a partial fraction decomposition.  Let,
\begin{equation*}
P = -(z^2 - z_+^2)(z^2 - z_-^2).
\end{equation*}
Then,
\begin{equation}
\label{eq:ground_state}
\begin{split}
\int_{z_-}^{z_+} \pi - \phi(z)\,dz &= - z_+ \E(k) + \left(1 - \frac{2E}{\Lambda}\right)\frac{1}{z_+}\times \\
 &\quad \left(\K(k) - \frac{1}{1-z_+^2}\Pi\left(\alpha^2, k\right)\right),
 \end{split}
\end{equation}
where $\K(k)$, $\E(k)$, and $\Pi(\alpha^2, k)$ are complete elliptic integrals of the first, second and third kinds, $k$ is the measure of orbit size defined in Eq.~\refeq{eq:df:k}, and
\begin{equation*}
\alpha^2 = \frac{z_+^2-z_-^2}{z_+^2 - 1}.
\end{equation*}

\subsection{Tunneling action $S_\epsilon$}

The ``tunneling action'' is defined analogously to the orbit action,
\begin{equation*}
S_\epsilon(E,\Lambda) = - \frac{N+1}{4\pi}\cdot 2\int_{-z_-(E,\Lambda)}^{z_-(E,\Lambda)} \abs{\pi-\phi(z,E,\Lambda)}\,dz,
\end{equation*}
with the absolute value necessary because $\phi(z,E,\Lambda)$ may be complex within the region of integration.  In fact, in the self-trapping region ($E > 1$, $\Lambda > 1$) the argument of the arccosine in $\phi(z,E,\Lambda)$ is smaller than $-1$ for all $z \in [-z_-, z_-]$.  Consequently, taking advantage of the identity,
\begin{equation*}
\arccos (-1-x) = \pi - \imath\, \arccosh (1+x),
\end{equation*}
one may rewrite $S_\epsilon$ as,
\begin{equation*}
S_\epsilon = -\frac{N+1}{\pi} \int_0^{z_-} \arccosh \left(\frac{2E - \Lambda z^2}{2\sqrt{1-z^2}}\right)\,dz.
\end{equation*}
As in the case of the orbit action, $S_\epsilon$ can be recast as an elliptic integral through integration by parts, and then reduced to a sum of canonical elliptic integrals using a partial fractions expansion.  The result is,
\begin{equation}
\label{eq:tunneling_action}
\begin{split}
-\frac{\pi S_\epsilon}{N+1} &= -\left(1 - \frac{2E}{\Lambda}\right) \frac{1}{z_+} \Pi(z_+^{-2}, k')\\
 &\quad + z_+ \left(\K(k') - \E(k')\right),
\end{split}
\end{equation}
where $k'=\sqrt{1-k^2}$ and we have used identity 19.6.5 in~\cite{DLMF}.

\section{Approximate solution to the quantization problem for large $N$}
\label{sec:appendix_approximate_quantization}

In this section, we derive an approximate semiclassical expression for the splitting by expanding the integrals of the previous section in small orbit sizes, $k$, and energies, $e$.

\subsection{Approximate orbit frequency}

To lowest order,
\begin{equation}
\label{eq:df:omega}
\omega = \frac{2\pi}{T} = \frac{\pi\Lambda z_+}{2 K(k)} = \sqrt{\Lambda^2 - 1} + O(\sqrt{e}).
\end{equation}
A higher-order expansion is unnecessary because $\Delta E$ depends on $e$ primarily through the tunneling phase in the exponent.

\subsection{Energy of the highest-energy state}

\begin{figure}
\begin{center}
\includegraphics[width=0.9\columnwidth]{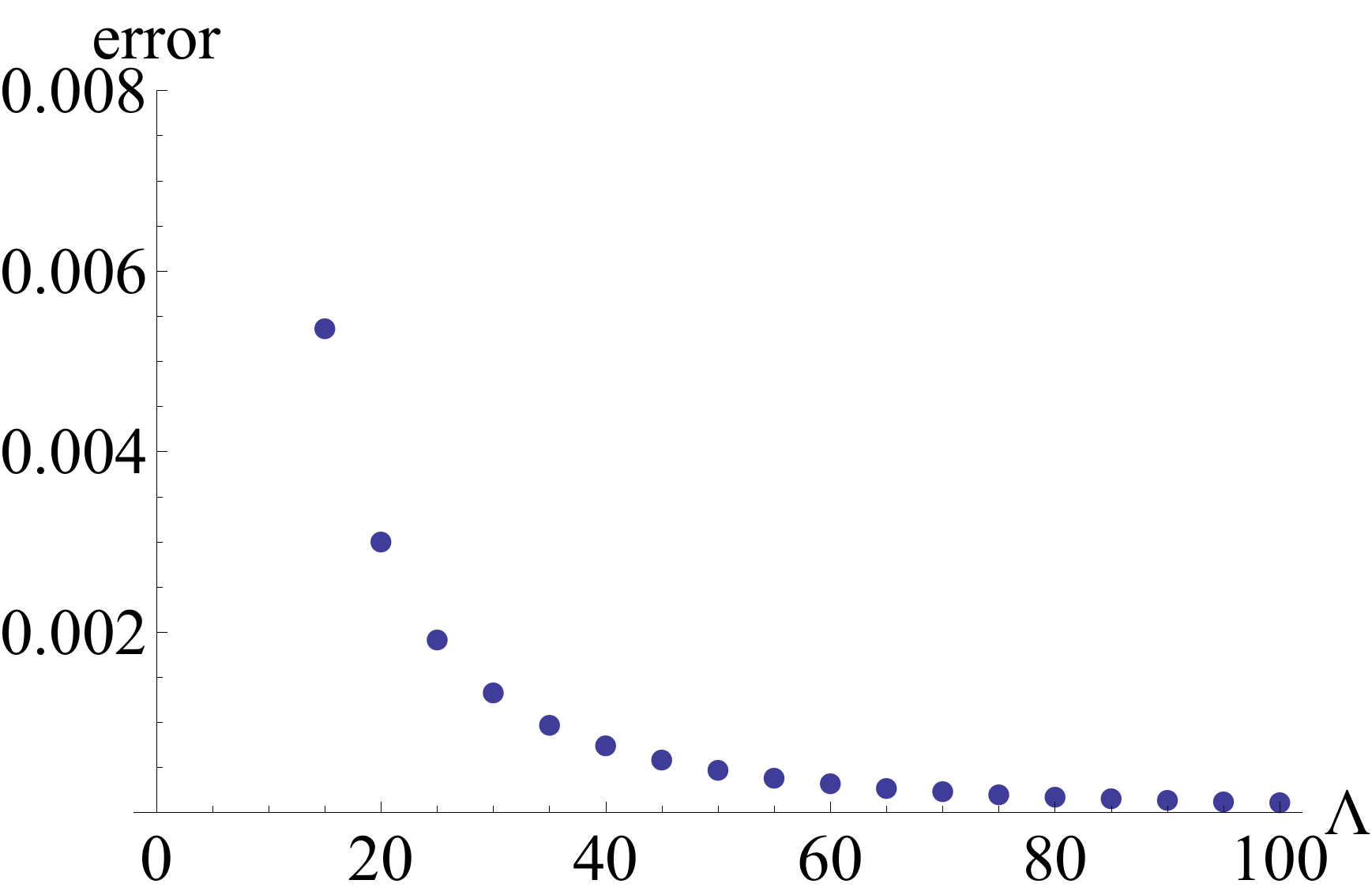}\\
\includegraphics[width=0.9\columnwidth]{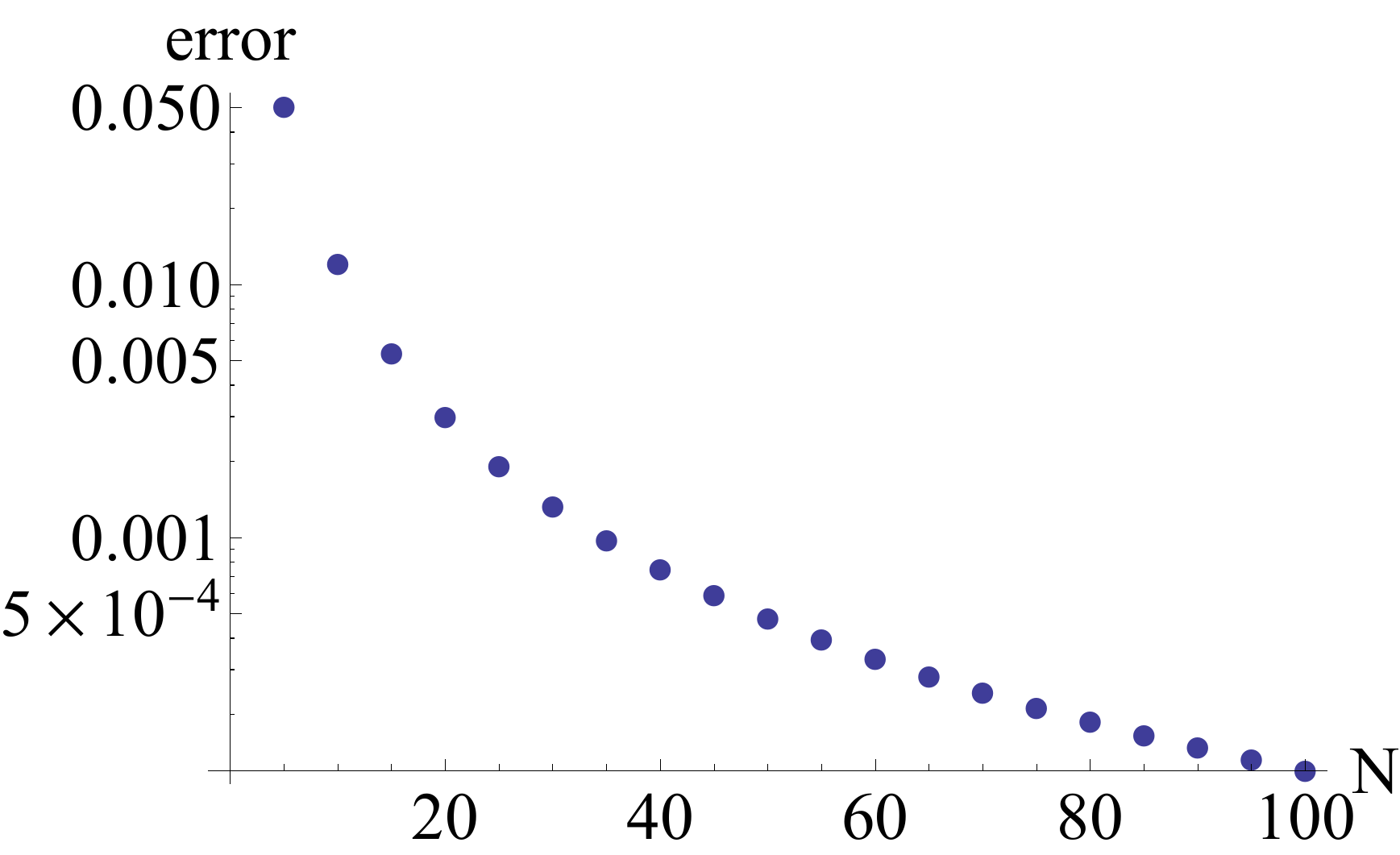}
\caption{Relative error in approximating the (numerically exact) solution of Eq.~\refeq{eq:quantization_condition_e} with the lowest-order approximation of Eq.~\refeq{eq:appendix_approx_ground_state}.  The figure on the left shows the dependence on $\Lambda$ (for $N = 20$) and that on the right---the dependence on $N$ (for $\Lambda = 2$). \label{fig:approx_ground_state}}
\end{center}
\end{figure}

Many of the quantities encountered in our discussion so far can be expressed more simply in terms of $e$ [the normalized energy relative to the maximum of $E$---see Eq.~\refeq{eq:df:e}] than $E$.  For instance, the classical turning points are,
\begin{equation*}
z_\pm = 1 - \frac{1}{\Lambda^2}\left(1 \mp (\Lambda-1)\sqrt{e}\right)^2
\end{equation*}
and the dimensionless measure of orbit size is,
\begin{equation*}
k^2 = \frac{z_+^2 - z_-^2}{z_+^2} = \frac{4\sqrt{e}}{(\sqrt{e}+1)^2 + \Lambda(1-e)}.
\end{equation*}
The quantization condition of Eq.~\refeq{eq:quantization_condition} reads,
\begin{multline}
\label{eq:quantization_condition_e}
\frac{\pi}{N+1} - \pi(1-z_+)\cdot\indicator{e>(\Lambda-1)^{-2}} = - z_+ \E(k)\\
 -\frac{1-(\Lambda-1)^2 e}{\Lambda^2}\frac{1}{z_+}\left(\K(k) - \frac{1}{1-z_+^2}\Pi(\alpha^2, k)\right),
\end{multline}
with $k$ and $z_\pm$ given by the expressions in the previous section.  Consider the case $e < (\Lambda-1)^{-2}$, when the highest-energy state orbit is a libration.  Expanding the elliptic integrals to lowest order in $k$ and then to lowest order in $e$ \footnote{Because the prefactors themselves depend on $e$, a consistent expansion requires expanding $\Pi$ to order $k^4$, though $\E$ and $\K$ are only expanded to order $k^2$.  (No odd powers of $k$ appear in the expansions.)} and then solving for $e$ gives a first-order estimate of the highest-energy state energy,
\begin{equation}
\label{eq:appendix_approx_ground_state}
e = \frac{2\Lambda \sqrt{\Lambda^2 - 1}}{(\Lambda - 1)^2 (N+1)}.
\end{equation}
As was already remarked in the main text, this estimate is very good.  See also Figure~\ref{fig:approx_ground_state}.

What happens if the nonlinearity is sufficiently high that the highest-energy state orbit is a rotation (i.e., $(\Lambda-1)^{-2} < e \ll 1$)?  It turns out that this case cannot be successfully treated using the same approach.  The term $\frac{1}{1-z_+^2}\Pi(\alpha^2, k)$ becomes ill-behaved, with both the prefactor and $\alpha^2$ very large.  The terms of the small-$k^2$ expansion of $\Pi(\alpha^2, k)$ are proportional to powers of $\alpha^2$ [cf.~Eq.~19.5.4 in~\cite{DLMF}], so keeping only the lowest-order terms in $k^2$ is no longer legitimate.  But the difficulty of extending our semiclassical method to this part of the parameter space is not a major concern, for two reasons:
\begin{enumerate}
\item The nonlinearity required for the ground state orbit to enclose the point $z = 1$ is large indeed, especially for larger atom numbers.  From Eq.~\refeq{eq:approx_ground_state}, the condition $e > (\Lambda -1 )^2$ can be estimated to imply,
\begin{equation}
2\Lambda \sqrt{\Lambda^2 - 1} \approx 2\Lambda^2 > N+1.
\end{equation}
\item The limit of very strong nonlinearity is particularly easy to treat using quantum perturbation theory~\citep{Bernstein1990, Dounas-Frazer2007,Salgueiro2007,Pudlik2013}.
\end{enumerate}

\subsection{Approximate tunneling action}

Finding a good large-$N$ approximation for the tunneling action [Eq.~\refeq{eq:tunneling_action}] is more difficult because both $\Pi(z_+^{-2}, k')$ and $\K(k') - \E(k')$ diverge in the limit $k' = \sqrt{1-k^2} \to 1_-$.  The lowest order asymptotic approximation is of $O(e^0)$:
\begin{equation*}
- \frac{\pi S_\epsilon}{N+1} \approx -\frac{\sqrt{\Lambda^2 -1}}{\Lambda} + \ln \left(\Lambda + \sqrt{\Lambda^2 - 1}\right).
\end{equation*}
It is possible to derive higher-order approximations by combining the known asymptotic expansions of the complete elliptic integrals, but they are complex and disappointingly inaccurate, except for large $N$ and either very large or very small $\Lambda$.

Instead of pursuing a formal expansion, let's attempt an \emph{ad hoc} improvement of the zeroth-order expression.  $S_\epsilon$ is a measure of the barrier to tunneling; as the ground state approaches the separatrix ($e \to 1$), the barrier should disappear.  The simplest way to enforce this behavior is to multiply the $O(e^0)$ expression by $(1-e)$:
\begin{equation}
\label{eq:ad_hoc_tunneling}
- \frac{\pi S_\epsilon}{N+1} \approx \left(-\frac{\sqrt{\Lambda^2 -1}}{\Lambda} + \ln \left(\Lambda + \sqrt{\Lambda^2 - 1}\right)\right)(1-e).
\end{equation}
This ansatz works remarkably well; furthermore, unlike the asymptotic expansions which may be either smaller or larger than the true value, Eq.~\refeq{eq:ad_hoc_tunneling} gives an upper bound on the magnitude of $S_\epsilon$ for all $\Lambda$.

\subsection{Approximate splitting formula}

\begin{figure}
\begin{center}
\includegraphics[width=0.9\columnwidth]{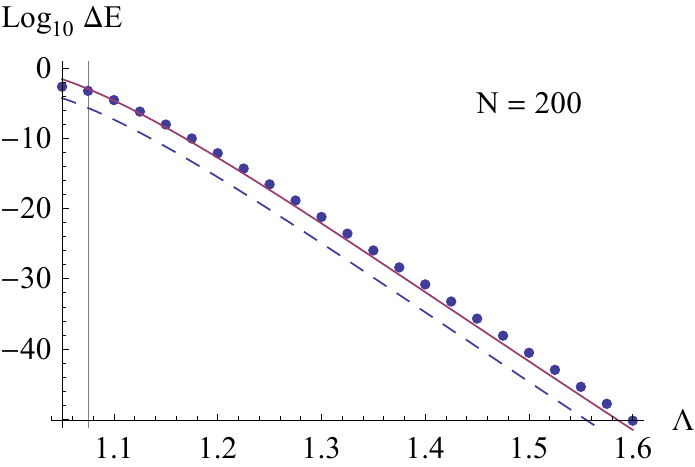}\\
\includegraphics[width=0.9\columnwidth]{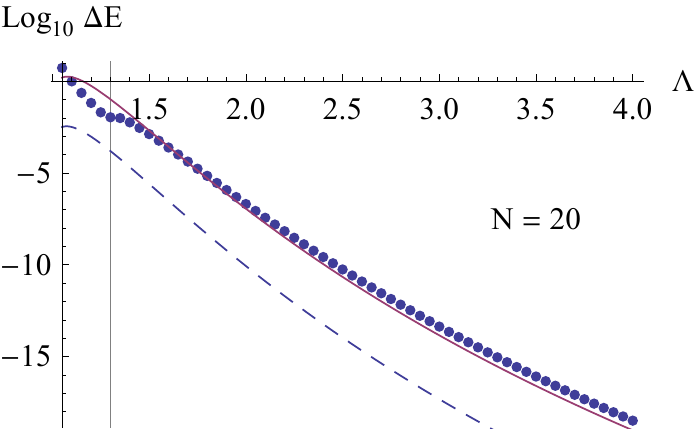}
\caption{Approximations to the semiclassical highest-energy state splitting.  Numerical solutions to Eq.~\refeq{eq:GraefeQuantization} are shows as blue dots; Eq.~\refeq{eq:approximate_splitting_3} is plotted as the solid red line, while Eq.~\refeq{eq:approximate_splitting_3} with $e=0$ is shown in dashed blue.  The black vertical line marks the point where the semiclassical approximation must break down because the area of phase space associated with the self-trapped region is less than $h/2$.
\label{fig:approx_splitting}}
\end{center}
\end{figure}

By combining the approximate expressions for the classical orbital frequency and the tunneling phase, we arrive at the following expression for the highest-energy state splitting:
\begin{equation*}
\Delta E \approx \frac{\hbar \omega}{\pi}\left(\frac{1}{\omega}\mathrm{e}^{-z_0}\right)^{(N+1)(1-e)},
\end{equation*}
where $z_0 \equiv \sqrt{1 - \frac{1}{\Lambda^2}}$ is the position of the classical potential maximum and $\omega = \Lambda \sqrt{1 - \frac{1}{\Lambda^2}}$ is the frequency of motion about it [cf. Eq.~\refeq{eq:df:omega}].  In this expression, the frequency is measured in the dimensionless units introduced with the Hamiltonian of Eq.~\refeq{eq:mean-field}.  In the units of $J$ and $U$ (Hertz),
\begin{equation}
\label{eq:approximate_splitting_3}
\Delta E \approx 2J\frac{\omega}{\pi}\left(\frac{1}{\omega}\mathrm{e}^{-z_0}\right)^{(N+1)(1-e)}.
\end{equation}

Figure~\ref{fig:approx_splitting} shows a comparison of this approximation with the numerical solution of the semiclassical quantization condition [Eq.~\refeq{eq:GraefeQuantization}].  Since our approximation to $S_\epsilon$ overestimates the barrier to tunneling, the tunneling frequency is generally underestimated, except close to the bifurcation where the dependence of $\omega$ on $e$ (which we neglect) becomes important.  Some qualitative features of the dependence of $\Delta E$ on $\Lambda$ can be reproduced even without the factor of $(1-e)$ in the exponent, and the agreement with the numerical solution improves as $N$ increases.  However, this $e=0$ approximation to $\Delta E$ is generally not within an order of magnitude of the numerically computed value.

\bibliography{bibliography}
\end{document}